# Observation of Zeeman effect in topological surface state with distinct material dependence

Ying-Shuang Fu[1,2*], T. Hanaguri[2 †], K. Igarashi[3], M. Kawamura[2], M. S. Bahramy[2,4], T. Sasagawa[3]

1. School of Physics and Wuhan National High Magnetic Field Center, Huazhong University of Science and Technology, Wuhan 430074, China

2. RIKEN Center for Emergent Matter Science, Wako, Saitama 351-0198, Japan

3. Materials and Structures Laboratory, Tokyo Institute of Technology, Yokohama, Kanagawa 226-8503, Japan

4. Department of Physics, University of Tokyo, Bunkyo-ku, Tokyo 113-0033, Japan

Email: *yfu@hust.edu.cn,  †hanaguri@riken.jp

**The helical Dirac fermions on the surface of topological insulators host novel relativistic quantum phenomena in solids [1,2]. Manipulating spins of topological surface state (TSS) represents an essential step towards exploring the theoretically predicted exotic states related to time reversal symmetry (TRS) breaking via magnetism or magnetic field [3-6]. Understanding Zeeman effect of TSS and determining its *g*-factor are pivotal for such manipulations in the latter form of TRS breaking. Here, we report those direct experimental observations in $Bi_2Se_3$ and $Sb_2Te_2Se$ by spectroscopic imaging scanning tunneling microscopy. The Zeeman shifting of zero mode Landau level is identified unambiguously by judiciously excluding the extrinsic influences associated with the non-linearity in the TSS band dispersion and the spatially varying potential. The *g*-factors of TSS in $Bi_2Se_3$ and $Sb_2Te_2Se$ are determined to be 18 and -6,**

**respectively. This remarkable material dependence opens a new route to control the spins in the TSS.**

When TRS of TSS is broken, a gap is opened at the Dirac point. This brings about novel topological excitations, such as magneto-electric effect [3,4], quantum anomalous Hall effect [3,5] and magnetic monopole effect [6]. Magnetic doping has proven to be an incisive way for TRS breaking via magnetic exchange interactions [7-9], thereby enabling the experimental observation of quantum anomalous Hall state [10]. However, magnetic dopants may introduce charge inhomogeneity [11] and weaken the spin-orbit coupling strength of the topological insulator (TI) compounds [12].

Zeeman effect, the coupling of spins with magnetic field, offers an alternative way for breaking TRS of TSS, which avoids the problem of magnetic doping. The opened gap is fully tunable by a perpendicular magnetic field $B$, i.e. $\Delta = g_s \mu_B B$, where $g_s$ is the electron $g$-factor of TSS and $\mu_B$ is a Bohr magneton [13]. Consequently, to manipulate the TSS via Zeeman effect, it is crucially necessary to know its $g$-factor.

Under $B$, Landau levels (LLs) are formed as a result of cyclotron motion of electron orbits. Meanwhile, Zeeman coupling to spins of electron influences the LL energies. Because spin degeneracy is lacking in the TSS, Zeeman effect causes energy shifting of LLs in $B$ (Fig. 1a), rather than splitting as observed in graphene [14] and conventional 2-dimensional (2D) electron systems [15]. The most pronounced Zeeman shift occurs to the zeroth LL, and decreases dramatically with increasing Landau index $n$ (Supplementary Section 1.1).

The Zeeman shifting behavior and *g*-factor determination of TSS have entailed intense research investigations primarily by quantum oscillation measurements [13,16,17]. However, the obtained results are still largely controversial. The inconsistency comes from actual materials being inevitably doped by defects. This prevents the lower LLs, which exhibit prominent Zeeman shift, from reaching Fermi level and contributing to quantum oscillation signals [16]. Further, the sign of $g_s$ cannot be determined from the Zeeman shift of nonzero LLs [17]. A recent tunneling spectroscopy study could probe $LL_0$ of the TSS formed at an interface with a conventional semiconductor [18]. However, the method depends on a band-bending of specific heterostructures, which cannot be readily applied to other TI compounds.

Spectroscopic imaging scanning tunneling microscopy (SI-STM) can access electronic states in a wide energy range with high spatial and energy resolution, rendering it feasible to study any LLs regardless of the doping level of TI compounds. In principle, Zeeman shift of $LL_0$ energy ($E_0$) of TSS is linear with *B*, and its slope determines $g_s$. Practically, however, more complicated factors hinder its direct observation. For one thing, a finite curvature is superimposed on the linear dispersion of TSS in actual compounds (Fig. 1b). This induces an extra *B*-linear change in $E_0$ that is irrelevant to Zeeman effect. For another, there exist spatial potential variations in TSS coming from the inhomogenously distributed charged defects [19]. This introduces an extrinsic *B* dependence of LL energies as the spatial extension of LL wave functions shrinks with increasing *B* (Fig. 1c).

To quantify the effects of non-linear dispersion and the potential variations, we consider the following model Hamiltonian [17].

$$H = \frac{1}{2m^*m_e}(\Pi_x^2 + \Pi_y^2) + v(\sigma_x\Pi_y - \sigma_y\Pi_x) + \frac{1}{2}g_s\mu_B B\sigma_z + V(x,y). \qquad (1)$$

Here $\vec{\sigma}$ are the Pauli matrices, $\vec{\Pi}$ are the canonical momentums. The first term depicts the nonlinearity of band dispersion, with $m^*$ as the effective mass relative to that of free electron ($m_e$). The second term depicts the helical Dirac fermions of TSS, with $v$ as electron velocity. The third term is the Zeeman term. The last term represents the potential variation. Energy of LL$_n$, $E_n$, without the last term of Eq. 1 has been given in Ref. 20.

For the effect of potential variation, we implement a 2D parabolic potential model $V(x,y) = E_D + \alpha_x x^2 + \alpha_y y^2$ to approximate the shape and location of potential extremes, where $E_D$ is Dirac-point energy. At the potential extreme, the $B$-dependent $E_0$ can be calculated using a first order approximation [21] (Supplementary Section 1.2). As a result, the $E_0(B)$ including the effects of non-linear dispersion and potential variations is given as

$$E_0 = E_D + \frac{1}{2}(\frac{2}{m^*} - g_s)\mu_B B + (\alpha_x + \alpha_y)\frac{\hbar}{|e|B}. \qquad (2)$$

Note that $m^*$ renormalizes $g_s$, and the potential variation introduces additional 1/$B$ dependence. Aiming to determine the intrinsic $g_s$, we have developed an analysis scheme of SI-STM data to correct these extrinsic factors, $m^*$ and $\alpha_x + \alpha_y$, and applied it to two different TI materials, $Bi_2Se_3$ and $Sb_2Te_2Se$.

First we evaluate the $m^*$ from the momentum-resolved LL spectroscopy [22]. Fig. 2a and b represent LL spectra of $Bi_2Se_3$ and $Sb_2Te_2Se$, respectively, under different $B$ measured at a fixed location of the two samples. In contrast to electron-doped $Bi_2Se_3$, defects in $Sb_2Te_2Se$ are acceptors locating its Dirac point in the empty state. Their $E_n$ exhibits a quasi-linear scaling relation with a scaling variable $(nB)^{1/2}$ (Figure 2c), which represents the energy-momentum dispersion of the TSS [22]. Since potential effect affects more to $E_n$ of small $n$ at high $B$ than that of large $n$ at low $B$ [21], the scaling with $(nB)^{1/2}$ demonstrates its influence is negligible at the measured location. Evidently, a finite curvature is dressed in the dispersion of both compounds indicating $m^*$ is finite, as is also captured by APRES measurements and band calculations [7,23]. Remarkably, the curvature of both compounds is very similar despite their different constituent elements. We evaluate the $m^*$ from the scaling function ignoring potential variations (Supplementary Section 3). For each compound, we did three measurements at different samples and got a $m^*$ of $0.12 \pm 0.03$ for $Bi_2Se_3$ and $0.13 \pm 0.02$ for $Sb_2Te_2Se$, which substantiates the observed similar band curvature.

Next, we assess the impact of potential variations on $E_0(B)$. The spatial variation of $E_0$ represents the potential landscape, despite it is smeared out by the magnetic length $l_B$ [19,21]. We start with $Bi_2Se_3$ and map out the potential landscape by performing a spectroscopic imaging of $E_0$ at a high $B$ of 11 T where $l_B \sim 7.7$ nm (Fig. 3a and d). We focus near the potential extremes (Figs. 3a and d) and fit their shapes with the 2D parabolic potential model (Supplementary Section 4). After positioning the tip at the fitted potential center (Fig. 3a and d, cross point), $B$ dependence of $LL_0$ peak was measured (Fig. 3b and e). The measured $E_0$ are plotted in Fig. 3c and f (black dots). Intriguingly, $E_0$ at potential minimum ($\alpha_x > 0, \alpha_y > 0$) and maximum ($\alpha_x < 0, \alpha_y < 0$) both exhibit $\sim 1/B$ behaviors but shift toward opposite directions. This is exactly expected from Eq. (2) and directly highlights the influence of potential variations on $E_0(B)$.

To eliminate the potential effect, we estimate the last term of Eq. (2) using the fitted values of $\alpha_x$ and $\alpha_y$. After subtracting the estimated potential effect, we get almost $B$ independent $E_0$ in the high $B$ region at both potential minimum and maximum (red symbols in Figs. 3c and d). This validates the methodology, and indicates the $m^*$ effect and Zeeman effect exert opposite influence, cancelling each other accidentally. By further subtracting the contribution from the $m^*$, we obtain the genuine Zeeman shift (Fig. 3 c, f, blue dots). As a result, the $g_s$ in $Bi_2Se_3$ is determined as $+18 \pm 4$. This differs significantly from its bulk value which is obtained as $+32$ by magneto-transport and NMR measurements [24,25].

The same methodology is implemented to $Sb_2Te_2Se$ (Fig. 4). Distinct from $Bi_2Se_3$, $LL_0$ state measured at the potential minimum (Fig. 4b) exhibits a nonmonotonic $B$ dependence (Fig. 4c) and *increases* with increasing $B$ at high $B$. Given that the $m^*$ of both compounds is almost the same, this demonstrates $g_s$ in $Sb_2Te_2Se$ is very different to that of Bi$_2$Se$_3$. Indeed we evaluate the $g_s$ in Sb$_2$Te$_2$Se to be -6 $\pm$ 2. Despite their similar energy dispersion, even sign of $g_s$ in Sb$_2$Te$_2$Se and Bi$_2$Se$_3$ is different.

We rationalize the above results under the framework of $\vec{k} \cdot \vec{p}$ theory. For narrow gap semiconductors, the conduction bands are coupled to the spin-orbit-split valence bands through second-order perturbative coupling, which substantially enhances the orbital sector of the *g*-factor. As a result, the total *g*-factor of such electrons/holes like their effective masses can significantly deviate from that of free electrons in both magnitude and sign [26]. In the case of TIs, strong spin-orbit-coupling creates a symmetry-inverted band gap. Thus, the atomic orbital character of the wave function undergoes a strong variation in the vicinity of the inverted band edges [27]. Such a variation is also manifested in the orbital character of TSS around the Dirac point [28]. Because Sb$_2$Te$_2$Se and Bi$_2$Se$_3$ have different atomic elements, atomic orbital compositions in their wave functions are rather different, which might explain the significant difference of $g_s$. Note that considerable energy dependence of $g_s$ is expected, since the orbital character of wave function changes notably in these systems. Our measurement is merely around the Dirac point, which is directly relevant to the gap opening of TSS via *B*. In this regard, the *g*-factor of TSS may be different from that of the bulk since they are measured at different energies. Further theoretical investigations accounting those factors are needed to develop a general theory to describe $g_s$.

Given the significant material dependence of $g_s$, we envision an interesting possibility of *g*-factor tailoring to TSS by controlling the chemical composition of the chalcogenide TI materials in the form of solid solutions [29, 30]. This opens up a new knob in manipulating the TSS for its spin-related applications.

**Methods:**

Experiments were performed with a modified commercial UNISOKU low temperature STM at 4.4 K or 1.5 K. Magnetic field up to 12 T can be applied perpendicular to the sample surface. $Sb_2Te_2Se$ and $Bi_2Se_3$ crystals grown by a modified Bridgman technique were cleaved *in situ* under ultrahigh vacuum conditions at ~ 77 K. After cleaving, the crystal was transferred quickly to the low temperature STM for subsequent measurements. Two $Bi_2Se_3$ and three $Sb_2Te_2Se$ samples were measured. A tungsten tip was used as STM probe which has been cleaned and characterized with a field-ion microscope. Tunneling spectra were obtained by lock-in detection of the tunneling current with a modulation voltage at 617.3 Hz feeding into the sample bias. The tip is grounded as the reference voltage.

**Acknowledgement:**

The authors thank Y. Tokura, X.-C. Xie, J.-H. Gao, P. Zhang, Z.-G. Wang, Y.-Y. Wang, X. Liu for helpful discussions. This work is funded by Grant-in-Aid for Scientific Research from the Ministry of Education, Culture, Sports, Science and Technology of Japan. Y.S. Fu acknowledges additional support from National Science Foundation of China.

**Author contributions:**

Y-S.F. and T.H. carried out the experiments and analyzed the data. K.I. and T.S. grew single crystals of $Sb_2Te_2Se$ and $Bi_2Se_3$. M.K. and M.S.B. supported theoretical understanding. T.H. supervised the project. Y-S.F. and T.H. wrote the manuscript with contributions from M.K. and M.S.B..

**Competing financial interests**

The authors declare no competing financial interests.

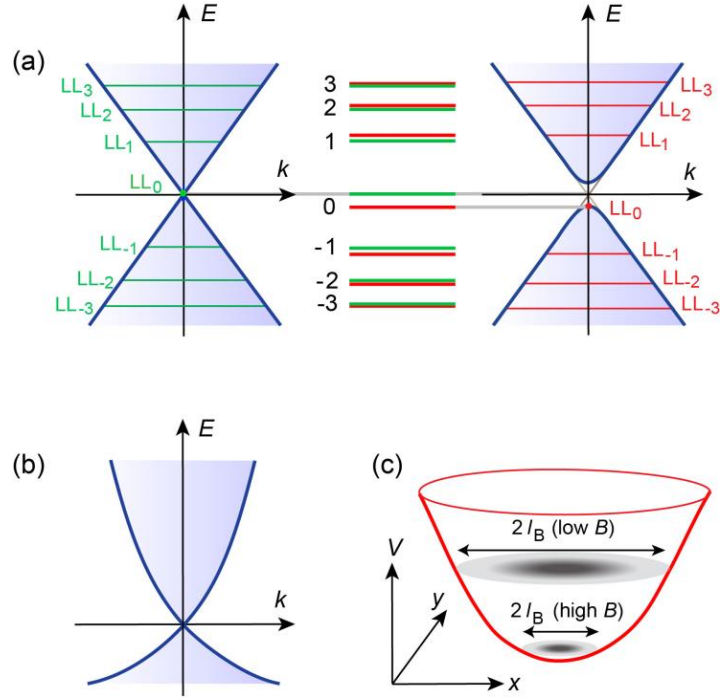

Fig. 1 Zeeman effect of TSS and extrinsic influences on its observation. (a). Schematics representing the Zeeman effect on LLs of TSS. When Zeeman effect is absent (left), LLs (green lines) are formed in a perpendicular $B$. When Zeeman effect is present (right), TSS becomes massive and its LLs (red lines) exhibit additional energy shift away from Dirac point. The amount of Zeeman shift decreases rapidly with increasing Landau index $n$. (b). Schematic showing band structure of actual TSS with a finite curvature superimposed on its linear dispersion. (c). Schematic of a 2D potential minimum and the spatial extension of $LL_0$ wave function at different $B$. Dark (light) grey color depicts high (low) intensity of the $LL_0$ wave function.

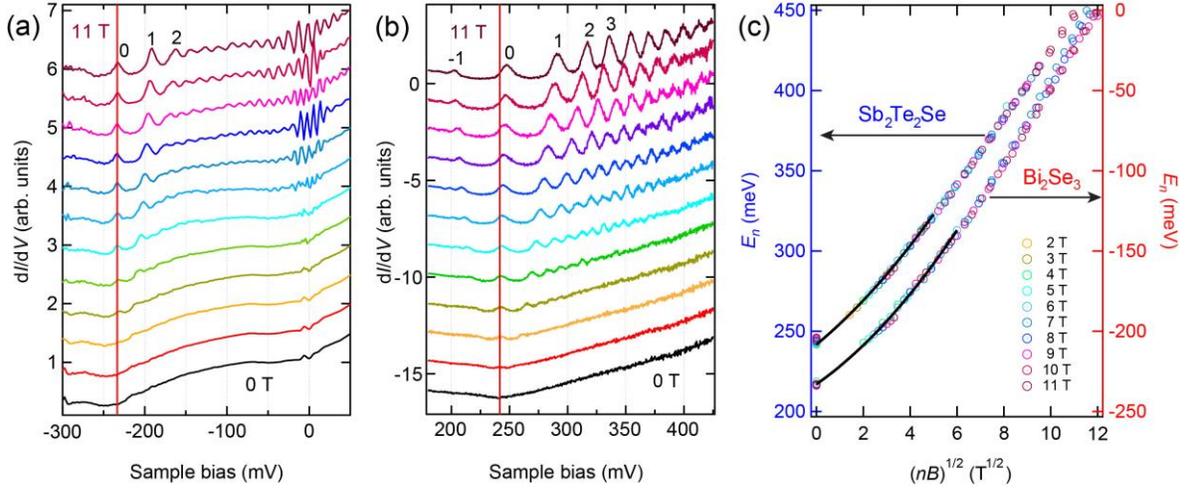

Fig. 2 LL spectroscopy of $Bi_2Se_3$ and $Sb_2Te_2Se$. Tunneling spectroscopy showing LLs of TSS measured at a fixed location of $Bi_2Se_3$ (a) and $Sb_2Te_2Se$ (b) surface at 1.5 K. The spectra were taken at different $B$ from 0 T to 11 T with an interval of 1 T, and are offset vertically for clarity. The red line marks energy of Dirac point. The spectra in (a) are taken from Ref. [22]. Measurement conditions of (b) were $V_s$ = -100 mV, $I_t$ = 50 pA, $V_{mod}$ = 1.4 $mV_{rms}$. (c). Scaling analysis of $E_n(B)$ based on data of $Sb_2Te_2Se$ in (b) and a comparison with that of $Bi_2Se_3$. $E_n$ are obtained by fitting the LL spectra with multiple Lorentz functions. Black curves depict fitting to the low energy parts of the scaling relations.

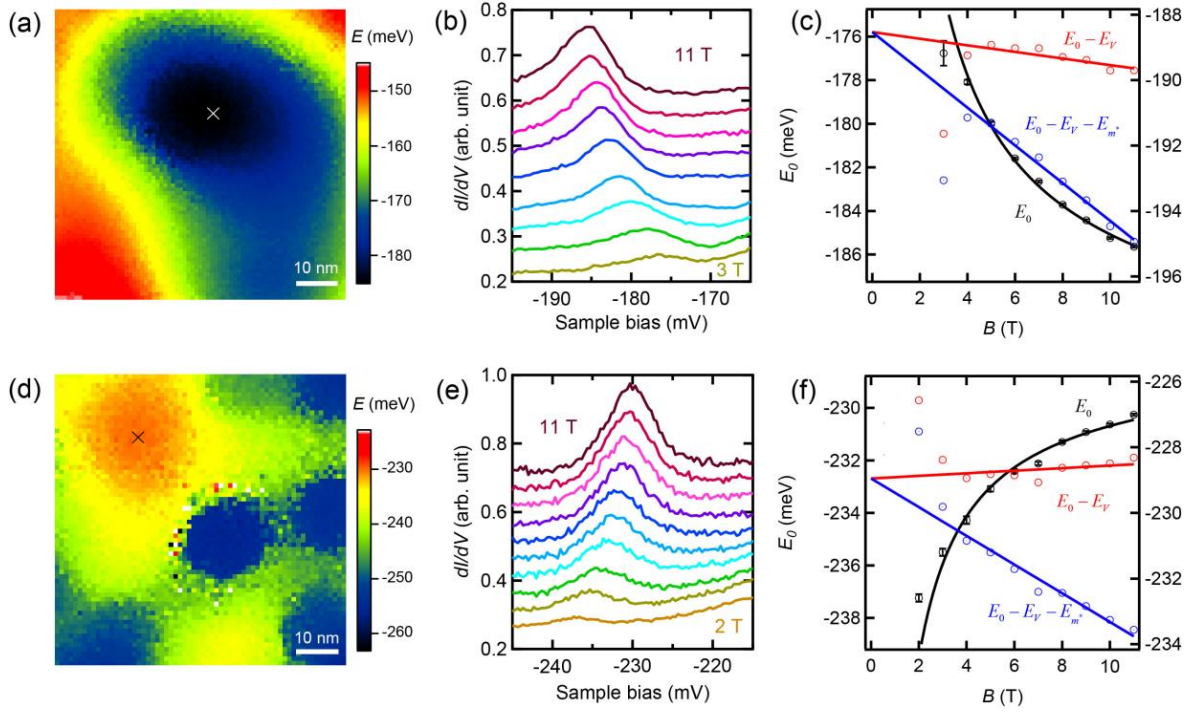

Fig. 3 Surface *g*-factor measurement on $Bi_2Se_3$. (a). Potential landscape of $Bi_2Se_3$ obtained by mapping $E_0$ at 11 T showing a potential minimum. Center of the potential minimum was determined by a 2D parabolic potential fitting and marked as a cross. Measurement conditions: $V_s$ = 50 mV, $I_t$ = 50 pA and $V_{mod}$ = 2.8 $mV_{rms}$, $T$ = 1.5 K. (b). Tunneling spectra taken at the potential minimum center in different $B$ from 3 T to 11 T with 1 T interval. The spectra have been shifted for clarity. Measurement conditions: $V_s$ = -220 mV, $I_t$ = 100 pA and $V_{mod}$ = 1.4 $mV_{rms}$, $T$ = 1.5 K. (c). $E_0$ at different $B$ is obtained by fitting data of (b) with a Lorentz line shape and plotted with black dots (left axis). The error bars are the standard deviation of fitting analysis. For comparison, $LL_0$ energies subtracting the effect of potential ($E_0 - E_V$) and further subtracting the effect of non-ideal dispersion ($E_0 - E_V - E_{m^*}$) are plotted with red and blue dots, respectively (right axis). Black curve denotes fitting to
12

$E_0(B)$ according to Eq. 2. Red and blue lines are the linear fitting of $(E_0 - E_V)(B)$ and $(E_0 - E_V - E_{m^*})(B)$, respectively. (d-f) Similar data and analysis as (a-c) to a potential maximum on $Bi_2Se_3$. Measurement conditions of (d) and (e) are both set at: $V_s$ = -200 mV, $I_t$ = 165 pA and $V_{mod}$ = 1.8 mV$_{rms}$, $T$ = 1.5 K.

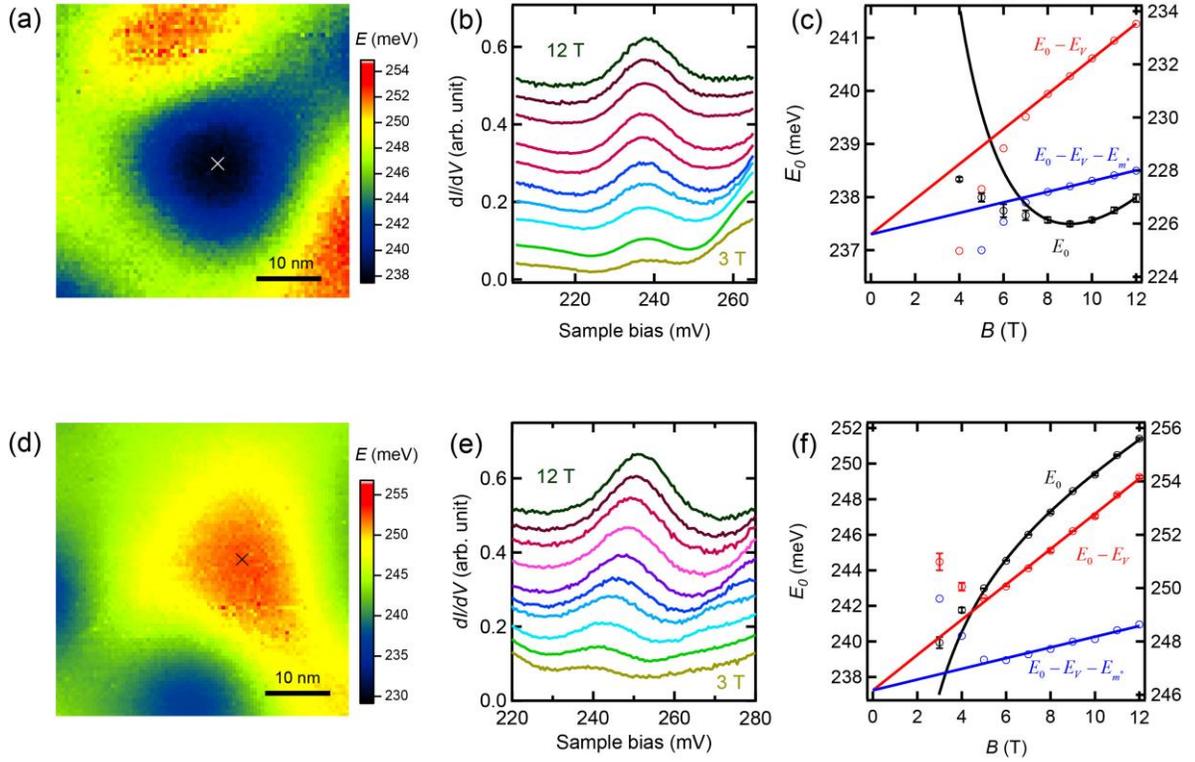

Fig. 4 Surface *g*-factor measurement on $Sb_2Te_2Se$. Similar data and analysis as Fig. 3 to a potential minimum (a-c) and a potential maximum (d-f) on $Sb_2Te_2Se$. Measurement conditions for (a) and (d): $V_s$ = 215 mV, $I_t$ = 50 pA and $V_{mod}$ = 2.8 mV$_{rms}$, $T$ = 4.4 K, $B$ = 12 T. Measurement conditions for (b) and (e): $V_s$ = 210 mV, $I_t$ = 50 pA and $V_{mod}$ = 1.8 mV$_{rms}$, $T$ = 4.4 K.

26. Roth, L.M., Lax, B. Zwerdling, S. Theroy of optical magneto-absorption effects in semiconductors. Phys. Rev. **114**, 90-104 (1959).

27. Liu, C.X., Qi, X.L., Zhang, H.J., Dai, X., Fang, Z., Zhang, S.C. Model Hamiltonian for topological insulators. Phys. Rev. B **82**, 045122 (2010).

28. Cao, Y. *et al.* Mapping the orbital wavefunction of the surface states in three-dimensional topological insulators, Nat. Phys. **9**, 499–504 (2013)

29. Arakane, T. *et al.* Tunable Dirac cone in the topological insulator $Bi_{2-x}Sb_xTe_{3-y}Se_y$. Nat. Commun. **3**, 636 (2012).

30. Zhang, J.S. *et al.* Band structure engineering in $Bi_{2-x}Sb_xTe_3$ topological insulators. Nat. Commun. **2**, 574 (2011).

Supplementary Information for

Observation of Zeeman effect in topological surface state with distinct material dependence

1. Models of LLs in TSS in the presence of Zeeman effect

We consider two models of TSS and study its LLs in the presence of Zeeman effect: one is ideal helical Dirac fermions, another one is non-ideal Dirac fermions perturbed by a parabolic curvature in their energy dispersion and potential variations.

1.1. Ideal helical Dirac fermions

The Hamiltonian for ideal helical Dirac fermions in a perpendicular magnetic field $B$ is given as:

$$H = v(\sigma_x \Pi_y - \sigma_y \Pi_x) + \frac{1}{2} g_s \mu_B B \sigma_z \quad (S1)$$

Here $\vec{\Pi} = \hbar \vec{k} + |e| \vec{A}$ is the canonical momentum with $\vec{k}$ and $\vec{A}$ being the momentum and the vector potential, respectively; $v$ is the velocity of electrons; $\sigma_i$ $(i = x, y, z)$ are Pauli matrices, and $g_s$ is the surface electron $g$-factor of TSS. We assume $B = |B|$ throughout the paper. After introducing ladder operators $a = \frac{l_B}{\sqrt{2}\hbar}(\Pi_y + i\Pi_x)$ and $a^\dagger = \frac{l_B}{\sqrt{2}\hbar}(\Pi_y - i\Pi_x)$, the Hamiltonian is written as

$$H = \frac{\sqrt{2}\hbar v}{l_B} \begin{pmatrix} 0 & a \\ a^\dagger & 0 \end{pmatrix} + \frac{1}{2} g_s \mu_B B \begin{pmatrix} 1 & 0 \\ 0 & -1 \end{pmatrix} \quad (S2)$$

Since the wave function is a 2-spinor $\psi_n = \begin{pmatrix} u_n \\ v_n \end{pmatrix}$, then we get

$$\frac{1}{2} g_s \mu_B B u_n + \frac{\sqrt{2}\hbar v}{l_B} a v_n = E_n u_n \quad (S3)$$

$$\frac{\sqrt{2}\hbar v}{l_B} a^\dagger u_n - \frac{1}{2} g_s \mu_B B v_n = E_n v_n \qquad (S4)$$

For ladder operators, $a^\dagger a |n\rangle = n|n\rangle$, $a^\dagger |n\rangle = \sqrt{n+1}|n+1\rangle$, $a|n\rangle = \sqrt{n}|n-1\rangle$, where $n$ is a non-negative integer. This in combination with Eq. S3 and S4 yields

$$E_{n\neq 0} = \pm\sqrt{2|e|\hbar v^2 nB + (g_s \mu_B B/2)^2} \qquad (S5)$$

$$\begin{cases} \psi^+_{n\neq 0} = \sqrt{\dfrac{1}{1+D_n^2}} \begin{pmatrix} D_n |n-1\rangle \\ |n\rangle \end{pmatrix} \\ \psi^-_{n\neq 0} = \sqrt{\dfrac{1}{1+D_n^2}} \begin{pmatrix} |n-1\rangle \\ -D_n |n\rangle \end{pmatrix} \end{cases} \qquad (S6)$$

where $D_n = \dfrac{1}{\sqrt{1+\dfrac{(g_s \mu_B B/2)^2}{2|e|\hbar v^2 nB}} - \dfrac{g_s \mu_B B/2}{\sqrt{2|e|\hbar v^2 nB}}}$. The positive (negative) brunch represents electrons (holes) of Dirac fermions.

Because $a|0\rangle = 0$, we get

$$E_{n=0} = -g_s \mu_B B/2, \text{ and } \psi_{n=0} = \begin{pmatrix} 0 \\ |0\rangle \end{pmatrix} \qquad (S7)$$

It is seen from Eq. S5 and S7 that the Zeeman shift of $LL_n$ can be estimated as $\Delta E_{n\neq 0} \cong \pm \dfrac{(g_s \mu_B B/2)^2}{\sqrt{2|e|\hbar v^2 nB}}$. Evidently, the largest shift occurs to $LL_0$ state and Zeeman shift of $LL_n$ dramatically decreases with increasing $n$. Energy-resolved spin magnetization is defined as $m_i = \dfrac{\hbar}{2}\langle \psi_n |\sigma_i| \psi_n \rangle$, $i = x, y, z$. Hence, the in-plane spin magnetization is zero.

And the out-of-plane spin magnetization is $m_{z,n=0}=-\dfrac{\hbar}{2}$ and $m_{z,n\neq 0}=\pm\dfrac{\hbar}{2}\cdot\dfrac{D_n^{\,2}-1}{D_n^{\,2}+1}$, which demonstrates $m_z$ decreases rapidly with increasing $n$.

We further evaluate the situation of reversing the direction of $B$. Applying negative $B$ is equivalent to probing the opposite surface of the topological insulator in positive $B$ [S1]. For negative $B$, the Hamiltonian becomes:

$$H = v(\sigma_x \Pi_y - \sigma_y \Pi_x) - \frac{1}{2} g_s \mu_B B \sigma_z \qquad (S8)$$

The ladder operators should accordingly change to $a = \dfrac{l_B}{\sqrt{2}\hbar}(\Pi_y - i\Pi_x)$ and $a^\dagger = \dfrac{l_B}{\sqrt{2}\hbar}(\Pi_y + i\Pi_x)$. Following similar algebra of positive $B$, we get the same expression of LL energies as Eq. S5 and S7. This is different from the conclusion from Ref. S2, which claims the Zeeman shift of $E_0$ is dependent on the applied direction of perpendicular $B$. We have experimentally justified this point by reversing the applied direction of perpendicular $B$. As shown in Fig. S1, the energies of LLs are the same for positive and negative $B$. The LL wave functions of negative $B$ have their upper and lower components switched compared to that of positive $B$. Therefore, its out-of-plane spin magnetization also reverses its sign.

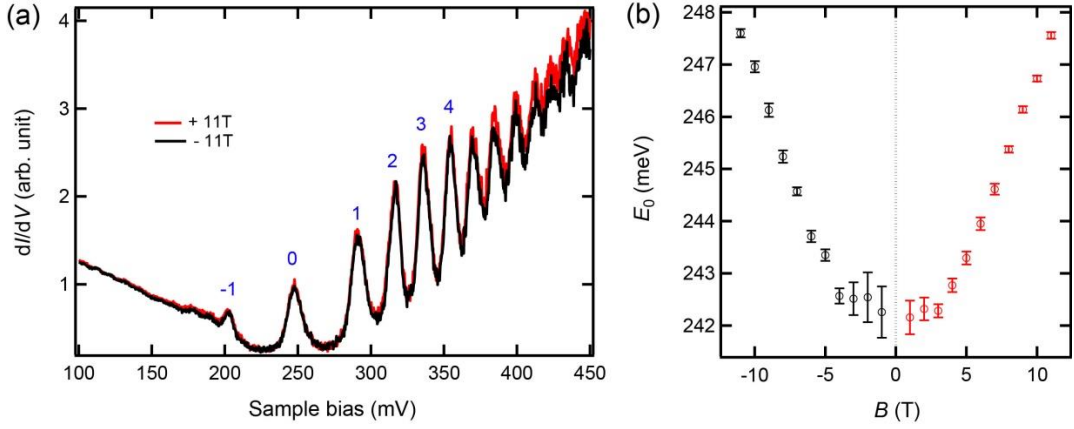

Fig. S1 LL spectra at positive and negative $B$. (a). STS spectra measured at the same location of $Sb_2Te_2Se$ showing LLs at positive and negative $B$ of 11 T. Measurement conditions are the same as Fig. 2b of main text. (b). $E_0$ value of $Sb_2Te_2Se$ extracted by fitting the $LL_0$ peak of (a) at positive (red dots) and negative (black dots) $B$ with a Lorentz shape. The errors are from the fitting.

**1.2. Non-ideal helical Dirac fermions**

Actual TSS is non-ideal because of the influence of finite curvature in its energy dispersion and potential variations in real space, which requires the modeling Hamiltonian to consider those factors. The Hamiltonian for non-ideal helical Dirac fermions having a parabolic term in its energy dispersion in a perpendicular $B$ is given as:

$$H = \frac{1}{2m^*m_e}(\Pi_x^2 + \Pi_y^2) + v(\sigma_x\Pi_y - \sigma_y\Pi_x) + \frac{1}{2}g_s\mu_B B\sigma_z \qquad (S9)$$

Where $m_e$ and $m^*$ are the absolute and relative effective mass of electrons, respectively. Its LL energies have been obtained in Ref. S3, which are written as:

$$E_{n\neq 0} = \hbar\omega_c n \pm \sqrt{2\hbar v^2|e|nB + (\frac{1}{2}\hbar\omega_c - \frac{1}{2}g_s\mu_B B)^2}$$

$$E_{n=0} = \frac{1}{2}\hbar\omega_c - \frac{1}{2}g_s\mu_B B = \frac{1}{2}(\frac{2}{m^*} - g_s)\mu_B B \qquad (S10)$$

Where $\omega_c = \dfrac{|e|B}{m^* m_e}$ is the cyclotron frequency of the parabolic electrons.

Next we model the influence of potential variations on the energy shift of LLs. We use a 2D parabolic potential model $V(x,y) = E_D + \alpha_x x^2 + \alpha_y y^2$ to approximate the potential extremes. At the potential extreme, energy shift of $E_0$ caused by potential variations to a first approximation is given analytically as [S4]:

$$E_V = \iint \phi_0 V \phi_0 \, dx\, dy = E_D + (\alpha_x + \alpha_y) l_B^2 = E_D + (\alpha_x + \alpha_y)\dfrac{\hbar}{|e|B} \qquad (S11)$$

Where $\phi_0 = \dfrac{1}{\sqrt{2\pi} l_B} \exp\left(-\dfrac{x^2 + y^2}{4 l_B^2}\right)$ is the wave function of $LL_0$ state. From Eq. S10 and S11, $E_0$ value accounting for Zeeman shift, finite $m^*$ and the potential variation can be given as Eq. 2 of main text.

## 2. Topography of Sb$_2$Te$_2$Se surface

Sb$_2$Te$_2$Se has tetradymite structure, which is the same as Bi$_2$Se$_3$. Its quintuple-layer unit consists of Te-Sb-Se-Sb-Te (Fig. S2, insert). Bonding forces between quintuple layers are weak van der Waals interactions. Therefore, the crystal can be easily cleaved. STM image of the cleaved surface clearly resolves the ordered atoms of triangular lattice (Fig. S2). Its lattice constant is estimated to be 4.2 Å, which is consistent with the bulk value. Since cleaving occurs between the adjacent Te layers, the imaged atoms should be Te.

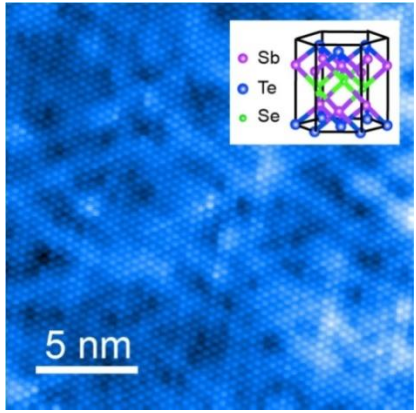

Fig. S2 Atomic resolution STM image showing the topography of $Sb_2Te_2Se$ surface. Imaging condition: $V_s$ = 250 mV, $I_t$ = 10 pA. Insert depicts the crystal structure of $Sb_2Te_2Se$.

3. **Evaluate the finite curvature of LL scaling**

We evaluate the $m^*$ that characterizes the finite curvature via the scaling analysis of LL energies in Fig. 2c of main text. Since only the Zeeman shift of $E_0$ is prominent, Eq. S10 can be approximately given as

$$E_{n\neq 0} \cong \frac{\hbar |e|}{m^* m_e} nB \pm \sqrt{2\hbar v^2 |e| nB} \ . \qquad (S12)$$

This indicates the scaling of $E_n$ with $(nB)^{1/2}$ still works even in the presence of the parabolic curvature. For $E_0$, its energy at 3 T is used for the scaling analysis, because its shift is negligible at low $B$. By fitting the low energy part of the scaling relation, $m^*$ value can be obtained. It is noted that the fitting at low energies gets worse considerably when high energies are also included. Since the effect of $m^*$ on $E_0$ should be determined by electronic states around the Dirac point, we only fit the low energy parts.

## 4. Fitting the potential extremes of $Bi_2Se_3$ and $Sb_2Te_2Se$ surface with 2D parabolic potential model

This section describes details of 2D parabolic fitting to the potential map. Fig. S3 a and d are potential maps of a potential minimum and a potential maximum on $Bi_2Se_3$ surface, respectively. They are the same as that of Fig. 3 in the main text, which are obtained by mapping $E_0$ at 11 T. We applied a 2D parabolic potential model to fit the potential extremes. We first draw two sectional lines across the potential extreme to estimate its shape and location. We then fit the sectional lines with a 1D parabolic potential. The delivered parameters are input as initial guess of the 2D parabolic fitting. The 2D fitting gives us the shape and location of the fitted potential, whose equipotential lines are superimposed on the potential map (Fig. S3 a and d). The 2D parabolic potential model can fit the measured potential map well, as can be inferred from the generated fitting error. To further evaluate the quality of fitting, we extract two sectional lines from the fitted potential (Fig. S3 b and c, red dots) and the potential map (Fig. S3 b and c, black curves). The sectional lines are along the major (line 1) and minor (line 2) axis of the fitted equipotential eclipse. As shown in Fig. S3 b and c (Fig. S3 e and f) that are extracted from Fig. S3a (Fig. S3d), the model fits to the measured data nicely. Similar analysis has been applied to the potential map of $Sb_2Te_2Se$ surface (Fig. S4).

Since the 2D parabolic potential applies to potential extremes only, it cannot be used any more at low $B$, when the spatial size of $LL_0$ state expands out of the potential extremes. Therefore, the fitting to relations of $E_0$ and $B$ according to Eq. 2 of main text only works above a certain critical field $B$ value ($B_c$). During fitting of black dots in Fig. 3c,f and Fig. 4c,f, we choose a lowest value of $B_c$, i.e. largest $B$-fitting region, without sacrificing the fitting quality. The size of $LL_0$ state ($2l_B$) at $B_c$ is depicted with a dashed circle in (a) and (d) of Fig. S3 and S4. The critical size of $LL_0$ state has considerable correspondence to the area of 2D parabolic potential fitting, which substantiates the model fitting.

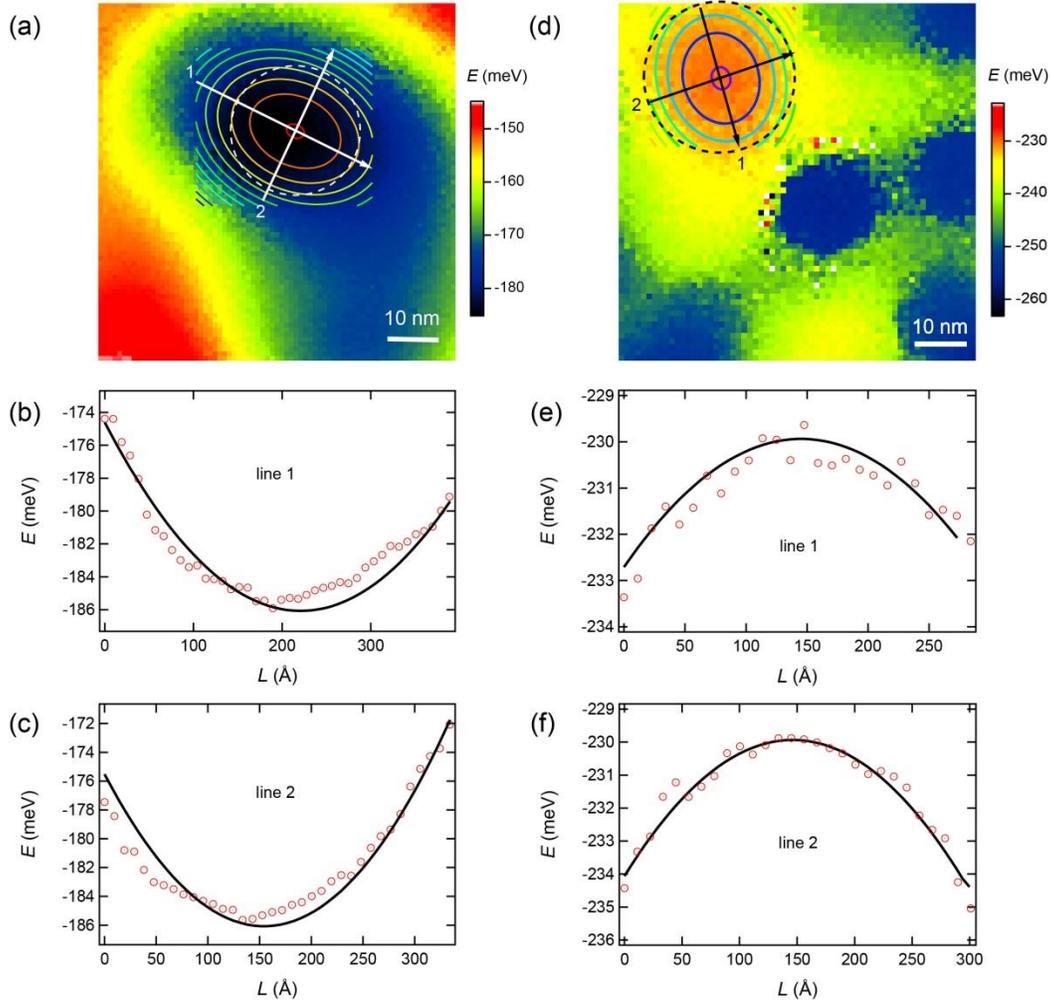

Fig. S3 Potential map of Bi$_2$Se$_3$ surface obtained by spectroscopic imaging $E_0$ at 11 T. The potential maps are the same as those Fig. 3 in main text. The potential extremes are fitted with a 2D parabolic potential model. Solid circles represent equipotential lines of the fitted potential. Adjacent potential lines have an energy interval of 2 meV for (a) and 1meV for (d). The most inner circle corresponds to -186 meV (-230 meV) for (a) [(d)]. Sectional lines are extracted from the $E_0$ map (red dots) and the fitting potential (black curves) along the major (line 1) and minor axis (line 2) of the fitted equipotential ellipse. The data in (b) and (c) are extracted from (a). The data in (e) and (f) are extracted from (d). Dashed circle in (a) [(d)] characterizes the location and size of LL$_0$ state at 4 T (3 T).

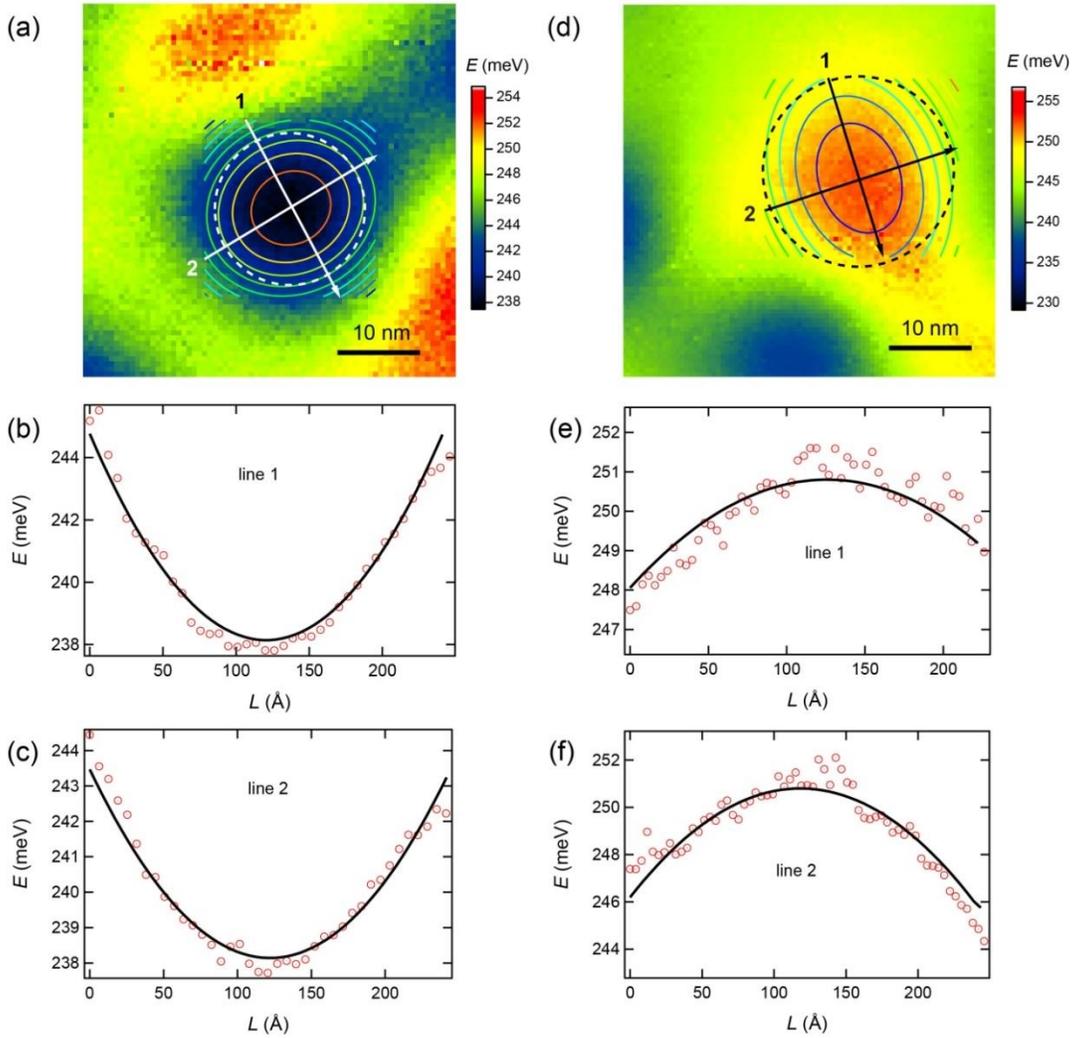

Fig. S4 Potential map of $Sb_2Te_2Se$ surface obtained by spectroscopic imaging $E_0$ at 12 T. The potential maps are the same as those of Fig. 4 in main text. The potential extremes are fitted with a 2D parabolic potential model. Solid circles represent equipotential lines of the fitted potential. Adjacent potential lines have an energy interval of 1meV. The most inner circle corresponds to 239 meV (250 meV) for (a) [(d)]. Sectional lines are extracted from the $E_0$ map (red dots) and the fitting potential (black curves) along the major (line 1) and minor axis (line 2) of the fitted equipotential ellipse. The data in (b) and (c) are extracted from (a). The data in (e) and (f) are extracted from (d). Dashed circle in (a) [(d)] characterizes the location and size of $LL_0$ state at 8 T (5 T).

## 5. Surface $g$-factor measurement on $Sb_2Te_2Se$ surface

In addition to Fig. 4 of the main text, we show the surface $g$-factor measured on 3 more potential extremes of $Sb_2Te_2Se$ surface. Fig. S5 a and d are the potential maps of two potential minimums and a potential maximum obtained by spectroscopic imaging $E_0$ at 12 T. Fitting the potential extremes with the 2D parabolic potential model delivers their shapes (Fig. S5f) and locations (marked as crosses in Fig. S5 a, d). We then focus on the potential centers and measure $E_0$ at different $B$ (Fig. S5 b, c, and e, black dots). The surface $g$-factor can be obtained by fitting the relation between $E_0$ and $B$ according to Eq. 2 of main text (Fig. S5 b, c, e, blue curves). Despite the potential extremes having different shapes, the extracted surface $g$-factors are all similar (Fig. S5f). The critical size of $LL_0$ state (dashed circles) also corresponds to the area of 2D parabolic fitting in Fig. S5 a,d.

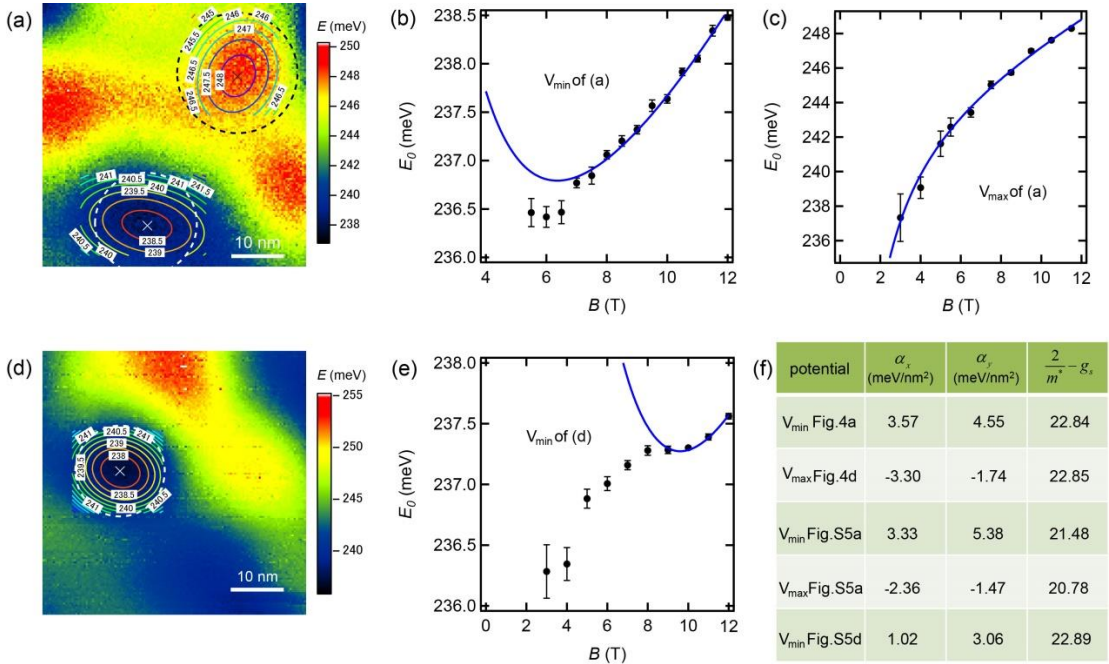

Fig. S5 Potential map of $Sb_2Te_2Se$ surface obtained by spectroscopic imaging $E_0$ at 12 T showing a potential minimum ($V_{min}$) and maximum ($V_{max}$) in (a) and a single potential

minimum in (d). Solid circles represent equipotential lines of the fitted 2D parabolic potential, whose energies are marked. Dashed circles in (a) and (d) characterize the location and size of $LL_0$ state at 7 T for $V_{min}$ of (a), 5 T for $V_{max}$ of (a), and 9 T for $V_{min}$ of (d). (b), (c) and (e) show $E_0$ at different $B$ (black dots) measured at fitted potential extremes (marked as crosses in a and d) and their fitting according to Eq. 2 of main text (blue curves). Error bars of $E_0$ are generated from the $LL_0$ peak fitting with Lorentz line shape. (f). Table showing fitting results of the shape of potential extremes and the surface $g$-factor including those shown in Fig. 4 of main text.

## 6. Modeling the effect of large potential extensions on $E_0$ at low $B$

In TSS of $Sb_2Te_2Se$, the effect of non-ideal dispersion and the Zeeman effect all tend to make $E_0$ shift to higher energy with increasing $B$. Meanwhile, the effect of a potential minimum tends to make $E_0$ shift to lower energy with increasing $B$. As a result, $E_0$ first decreases and then increases with $B$ for a potential minimum, as is seen from Fig. 4c. However, shifting trend of $E_0$ differs at different potential minimums. For instance, Fig. S5e exhibits a monotonic shift of $E_0$ with $B$. As the spatial extension of $LL_0$ state increases with decreasing $B$, the potential at large extensions gets involved and affects the shifting trend of $E_0$. In this section, we model its effect to understand the observed different shifting behavior of $E_0$ at different potential minimums.

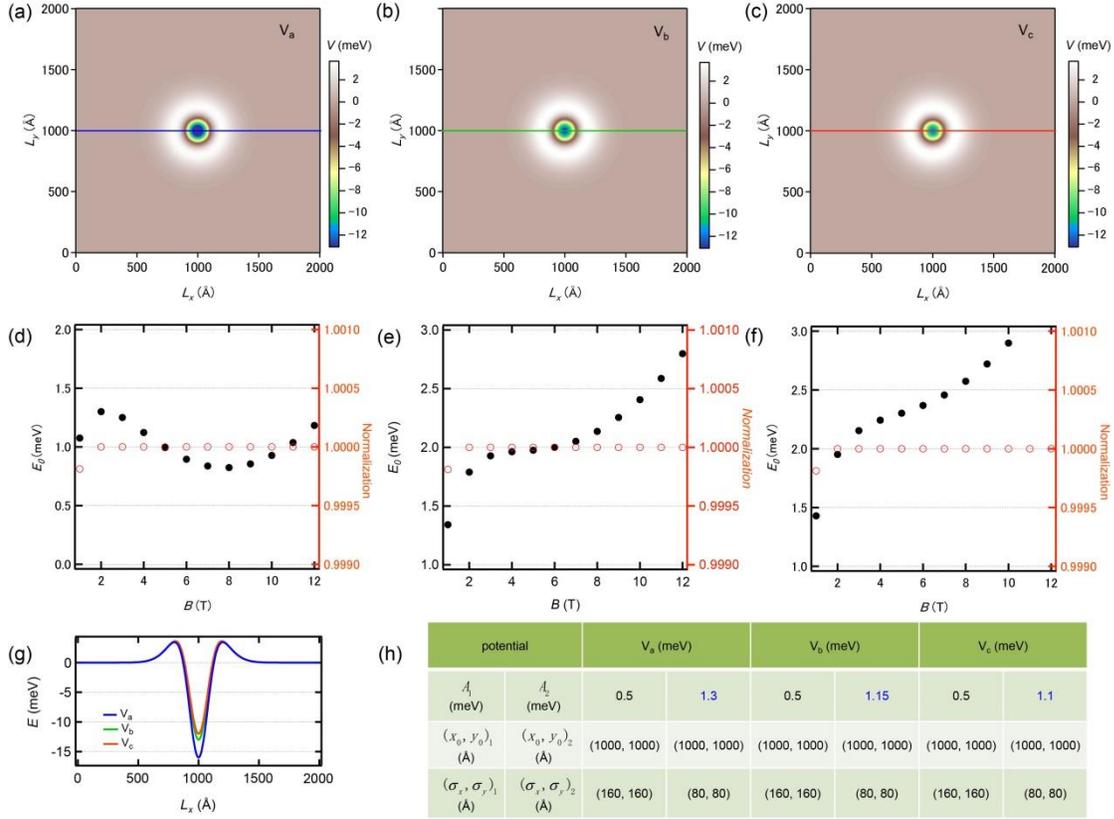

Fig. S6 Model calculation based on a single potential. (a-c) Single potentials with different shapes that are all a superposition of a Gaussian maximum and a Gaussian minimum. Pixel size: 512×512. (d-f) Calculated $E_0$ and the normalization value of $LL_0$ state at different $B$ according to Eq. S13 and S14. (g) Sectional lines of Potential a-c across the potential center (horizontal lines in a-c). (h) Table showing parameters of Potential a-c. Different parameters among the three potentials are highlighted with blue color.

We first simply use two superimposed Gaussian potentials to model the potential minimum, because they contain different types of variations and go flat at large spatial extensions. Gaussian potential has the form $V^G(x,y) = A\exp(-(\frac{(x-x_0)^2}{2\sigma_x^2} + \frac{(y-y_0)^2}{2\sigma_y^2}))$, where $A$ is the amplitude, $(x_0, y_0)$ is the center, $(\sigma_x, \sigma_y)$ is the decay length of potential. The model potential is thus $V(x,y) = V^G{}_1 - V^G{}_2$. We modeled 3 potentials (Fig. S6 a-c) to demonstrate the different shifting behavior of $E_0$ with $B$, and their parameters are listed in Fig. S6h. The 3 potentials all have a dip in the center, but decrease in energy as their spatial size gets extended, forming a hump shape. The depth of the potential dip is deepest for Potential a ($V_a$), and shallowest for Potential c ($V_c$) (Fig. S6g).

Subsequently, energy of $E_0$ can be calculated according to Eq. S10 and S11 as

$$E_0 = \iint \phi_0 V \phi_0 \, dxdy + \frac{1}{2}(\frac{2}{m^*} - g_s)\mu_B B \qquad (S13)$$

Where $\phi_0 = \frac{1}{\sqrt{2\pi}l_B}\exp(-\frac{x^2+y^2}{4l_B{}^2})$ is the wave function of $LL_0$. The first term depicts the potential effect, and the second term represents influence from the non-ideal dispersion and the Zeeman effect. We use a value of 20 for $\frac{2}{m^*} - g_s$, which is close to the experimentally measured value of 19. The integration is calculated in the range of $[x, y] = [(0, 2000), (0, 2000)]$. To guarantee the $LL_0$ state at all calculated $B$ is within the integration range, we always check its normalization value to be 1 or not (Fig. S6 d-f), which is given as

$$N_0 = \int_0^{2000}\int_0^{2000} |\phi_0|^2 \, dxdy \qquad (S14)$$

The calculated $E_0$ at the potential center exhibits different energy shifting trend with $B$ at different potential minimums (Fig. S6 d-f). When the potential dip is deep (Potential a), $LL_0$ state is mostly weighted by to the potential dip even at low $B$. Therefore, its $E_0$ first decreases and then increases with decreasing $B$ (Fig. S6d), which is expected for a potential minimum and is similar as Fig. 4c. It is noted the normalization value of $LL_0$ state at 1 T is less than 1, which means the calculated $E_0$ at 1 T should be neglected. When the potential dip is shallow (Potential b), weighting of $LL_0$ state at the potential hump is significantly enhanced. This makes $E_0$ shift monotonically with decreasing $B$ (Fig. S6e), which is similar as Fig. S5e. The monotonic shifting behavior of $E_0$ is more obvious (Fi. S6f) as the potential dip gets even shallower (Potential c).

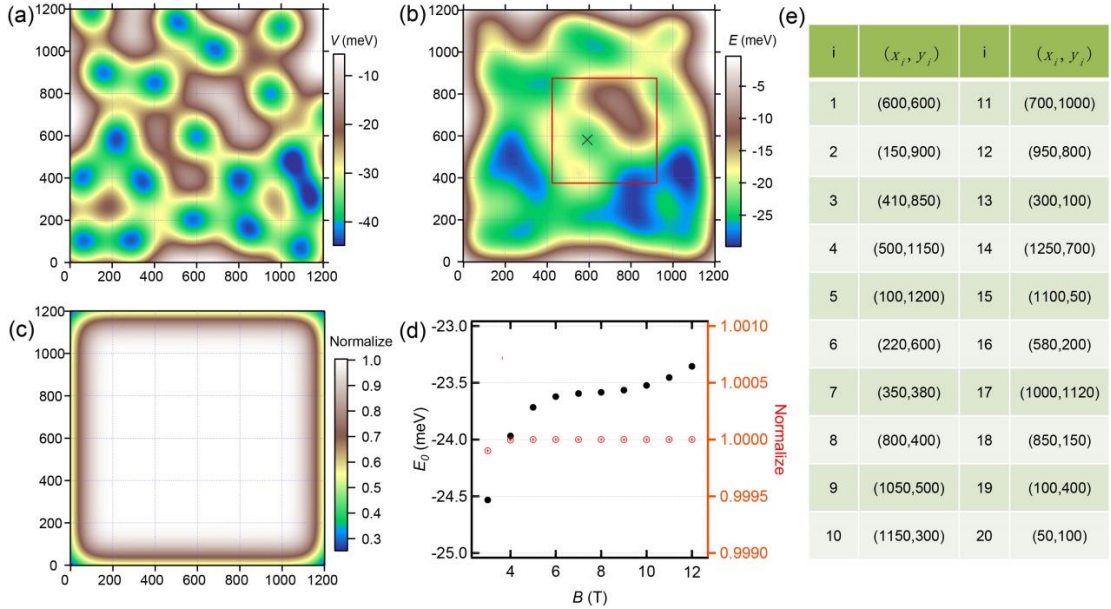

Fig. S7 Model calculation based on multi-potential minimums. (a-c) A multi-minimum potential composed of 20 identical Gaussian dips. (b) Calculated $E_0$ map at 12 T according to Eq. S13. A red rectangle corresponds to the modeled potential map of Fig. S5d. (c) Calculated normalization map of $LL_0$ state at 12 T. (d) Calculated $E_0$ and normalization of $LL_0$ state at the potential minimum center of the rectangle area (marked as cross in b) at different $B$. (g) Table showing parameters of Gaussian potential minimums for constructing the potential in (a). Pixel size of (a-c): 512×512. Value of $\frac{2}{m^*} - g_s$ is 20 for the calculation.

On the basis of a single potential, we further constructed a multi-minimum potential aiming to model the actual potential variations of Fig. S5d. Fig. S7a shows the modeled potential, which is composed of 20 identical Gaussian potential dips with different potential center coordinates, i.e.

$$V(x,y) = \sum_{i=1}^{20} V^G{}_i(x,y) \qquad (S15)$$

Where $V^G{}_i(x,y) = 2\exp(-(\frac{(x-x_i)^2}{2\times 100^2} + \frac{(y-y_i)^2}{2\times 100^2}))$ is the single Gaussian potential, and the parameters of $(x_i, y_i)$ are listed in Fig. S7e.

Then, the $E_0$ map at 12 T can be obtained by calculating energy of $E_0$ at every pixel point according to Eq. S13 (Fig. S7b). Since our integration range is limited to $[x, y] = [(0,1200), (0,1200)]$, we further calculate the normalization of $LL_0$ state at every pixel point (Fig. S7c), and those boundary regions whose normalization values smaller than 1 should be neglected. The calculated $E_0$ map (Fig. S7b, red rectangle) nicely reproduces the measurement of Fig. S5d. We then focus on the potential minimum (marked as cross in Fig. S7b), and calculate its $E_0$ at different $B$. The obtained result (Fig. S7d) also reproduces the shifting trend of Fig. 5e. Therefore, our model calculations substantiate our experimental observations, demonstrating that the potential at large extensions could not only affect the amount of shifting of $E_0$ but also change its trend with $B$.